\documentclass{kluwer}
\usepackage{epsfig}

\begin{document}
\begin{article}
\begin{opening}

\title{\bf The morphological evolution of galaxy satellites}

\author {Lucio \surname{Mayer$^1$}}

\author {Fabio \surname{Governato$^2$}}

\author {Monica \surname{Colpi$^1$}}

\institute {$^1$Dipartimento di Fisica Universita' Degli Studi di Milano,
Italy, $^2$ Osservatorio Astronomico di Brera, Merate, Italy}

\author {Ben \surname{Moore$^3$}}

\author {Thomas R. \surname {Quinn$^4$}}

\author {Carlton M. \surname {Baugh$^3$}}

\institute {$^3$Physics Department, University of Durham, Science Labs.,
Durham, U.K, $^4$Department of Astronomy, University of Washington, Seattle,
WA, USA} 

\begin{abstract}

We study the evolution of galaxy satellites with high resolution
N-body simulations.  Satellites are modeled as replicas of typical
low and high surface brightness galaxies (LSBs and HSBs).  Encounters on
high eccentricity orbits (as typical in hierarchical models of galaxy
formation) strip LSBs of most of their stars and tend to decrease
their surface brightness.  On the contrary, bar instability in HSBs
leads to substantial loss of angular momentum of the stellar component
and to an increase of central surface brightness.  In both cases the
remnant resembles a spheroidal galaxy with an exponential surface brightness
profile.
A simple modeling of color
evolution and interactions driven star formation gives M/L ratios for
the remnants that are roughly consistent with observations. These results
suggest an evolutionary scenario for the dwarf galaxies in our Local
Group, faint dSphs being the descendents of LSBs and  brighter dSphs/dEs
being the final state of HSB satellites.

\end{abstract}                    
\end{opening}

\section{Introduction}

Our knowledge of the galaxies of the Local Group is
becoming increasingly detailed: we have substantial information
regarding star formation histories, kinematics and morphology of many faint
dwarf satellites of the Milky Way and Andromeda (c.f. Mateo 1998 and
references within). 

The Local Group shows a  morphology density relation
that resembles that found in galaxy clusters (Dressler et al. 1998).
Dwarf irregulars (dIrrs) are 
found mainly in the far reaches of the Local
Group, while
dwarf spheroidals (dSphs) and dwarf ellipticals (dSphs/dEs)
are abundant close to the primary galaxies.  

Detection of tidal
streams in the halo of the Milky Way (de Zeeuw 1998) and peculiar
structure of some of its closest neighbors (Ibata \& Lewis 1998)
suggest that mutual interactions between galaxies  have played an
important role in the evolution of all Local Group members.

In this contribution, we explore the effect that tidal interactions
with the primary galaxies could have on accreting disk-like satellites.

\section{Galaxy models}    

Galaxy models were built using the method developed by Hernquist
(1993).
We used observational constraints as well as 
theoretical models of galaxy formation (Cole et al. 1999) 
to make credible replicas of real galaxy satellites.
Parameters for disk and halo components were chosen to represent
typical HSB and LSB dwarf galaxies.  We start by choosing a circular
velocity ($V_c \sim 75$ km/s), comparable to that of large companions of
spiral galaxies (Zaritsky et al. 1993), such as the LMC or NGC205 in our
Local Group.  The virial mass of the satellite is then determined by the
circular velocity and is weakly dependent on cosmology (White \& Frenk
1991)
(we assumed a CDM model with $\Omega=1$ and $h=0.5$).  
 
 HSBs and LSBs
 obey the same B-band Tully Fisher relation on a large range of
 circular velocities (Zwaan et al. 1997): using this last relation we
 derive a value for the disk luminosity $L{_B} \sim 2 \times 10^{9}
 L_{\odot}$.
 We assign the same disk mass to both HSBs and LSBs by assuming
 $M/L{_B} = 2$, as suggested by the generalized Bottema
 model (de Blok \& McGaugh  1997). 
 Disks are constructed using a Toomre parameter $Q=2$, this being 
 a necessary condition for global stability against bar modes
 (Friedli, these proceedings).
 The HSB disks have an exponential scale-length $r_{h}=2$ kpc while we
 use $r_{h}=5$ kpc for LSB disks: these values are consistent with
 the observed $V_{c}-r_{h}$ relation (Zwaan et al. 1997) and  give a
 surface brightness $\mu_{0} = 22$ mag arcsec$^{-2}$ for the HSB
 satellite and $\mu_{0} = 24$ mag arcsec$^{-2}$ for the LSB
 satellite. This is  in good  agreement with average values found in the samples of
 de Blok \& McGaugh (1997).
 Each galaxy model is embedded in an isothermal halo truncated at the virial
 radius (the same for both models as  it depends only on V$_c$)
: the halo is $60$ times more massive than the disk and has 
 a core radius $r_c$ equal
 to $r_{h}$.
 Due to its larger core,
 the LSB model has a low-concentration halo and thus a slowly rising
 rotation curve, while the HSB satellite has a $3$ times more concentrated
halo
 and a steeply rising rotation curve, consistent with observations (de
 Blok \& McGaugh 1997) (see Fig.1). $M_{total}/M_{disk}$ at the
 "optical radius" $R_{opt} = 3 r_{h}$ (see Persic and Salucci 1997)
 is equal to $4$ for the HSB and  is over $10$ for the LSB
 satellite.  We used about $200.000$ particles for the halo and
 $50.000$ particles for the disk of each satellite model.  

% Note that
% models built following the classic ``maximum disk'' hypothesis for
% the disk/halo decomposition would produce disks strongly unstable to
% bar formation.

Halo particles that pass through the disk are less massive (and hence
proportionally more numerous) than halo particles whose orbits do not
intersect the disk. This reduces disk heating 
due to two-body scattering by heavier halo particles (Lacey \& Ostriker
1985,
Velasquez \& White
1998).
The softenings are set to $0.06r_{h}$ for the disk,
$0.4r_{h}$ for the lo-res
halo particles and $0.35r_{h}$ for the hi-res ones.  The models were
evolved in isolation for more than 5 Gyr to test their stability.  At
a fixed particle number we verified that disk heating is reduced by a
factor of about 2 using a variable resolution model for the
satellites' halos.

The primary galaxy is modeled as a Milky-Way sized isothermal halo
 ($V_c \sim$ 220 km/s) truncated at the virial radius and whose mass is
 then 30 times larger than that of the satellites. It is represented
 by either a 50000 particles N-Body realization or simply by an
 external potential. This last configuration is preferable as it
 avoids numerical disk heating due to
 two-body scattering by massive particles belonging to the primary
 halo. Dynamical friction can be safely neglected because of the
 small mass of the satellites and the further delay resulting from tidal
 stripping (Colpi et al., in preparation).
 Our results are independent of the type of halo
 actually used.

\section{Initial Conditions}  

We perform all of our simulations with the parallel treecode PKDGRAV
(Stadel et al., in preparation) 
which has multistepping capabilities and uses
local acceleration for the assignment of individual timesteps to
particles. The minimum allowed timestep is $\sim 5 \times 10^{5} yr$.
Force calculations are done
using a multipole expansion up to hexadecapole terms with a tolerance
parameter $\theta = 0.7$.  The satellite is put on a bound
and very eccentric orbit (with $apo/peri=10$ or $4$) with an apocenter
close to the virial radius of the primary system
(consistent with the satellite just being accreted).  
%These satellites
%would not interact directly with the disk of the primary but their
%orbit would carry them relatively close to the central region of the
%primary. 
Orbits of this kind are the most common for satellite halos
in cosmological N-Body simulations (Ghigna et al. 1998).  We run
simulations with different relative orientations of the orbital
angular momentum and spin of the satellite, from pure prograde
encounters (vectors are parallel) to pure retrograde encounters
(vectors are antiparallel).  Several numerical tests (i.e varying
timestep, tolerance and running identical i.c. with the TREECODEV3 by
Barnes \& Hernquist) were performed to ensure that results do not
depend on the code or the particular choice of numerical parameters.

\section{Results}

\begin{figure}
\centerline{\epsfig{file=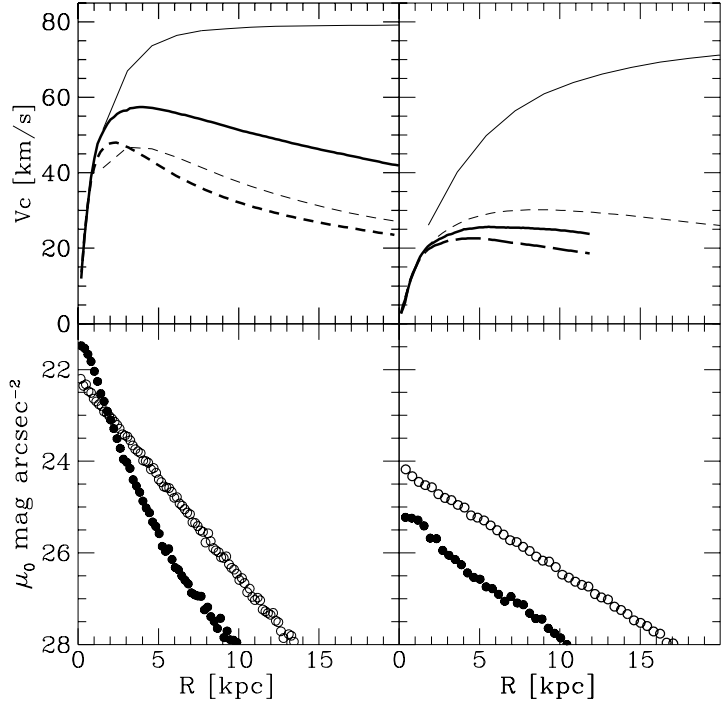,height =10cm}}
\caption{
Left: HSB satellite on an orbit with $apo/peri=10$.
Right: LSB satellite on the same orbit.
Upper  panel: Evolution of circular velocity profiles
$(V_{c}=\sqrt{(GM(r)/r)})$.
Thin lines represent total profile (solid) and stellar profile (dashed) at
$t=0$. Thick lines represent total (solid) and stars (dashed) after $t \sim 7$
Gyr.  Lower panels: Evolution of surface brightness profiles in the B band.
Open dots are for $t=0$, filled dots for $t \sim 7$ Gyr.}
\end{figure}

HSBs and LSBs lose most of their
dark matter halo after a few orbital periods (of typically 2 Gyr). 
The ratio $M_{total}/M_{stars}$ at the initial $R_{opt}$ 
has decreased to $1.5-4$ after $t \sim$ 7 Gyr (see Fig.1). 
LSBs are structurally very fragile compared to HSBs: 
they end up with  the smallest halos  and lose
up to $ 90 \% $ of their stellar mass, 
decreasing their total mass to $\sim 10^8 M_{\odot}$.
Instead, HSBs lose no more than $40 \%$ of their stellar mass
and have final total masses in excess of $10^{9} M_{\odot}$.
The different response of the satellites is due primarily to the potential 
depth of the mass distribution and the disk scale-lengths (e.g. Moore 1999).
A stellar bar appears after the first pericentric passage, its pattern
being particularly strong for HSBs as a consequence of the higher disk
surface density (Mihos et al. 1997). This leads to substantial angular momentum
loss for HSBs disk particles (see Fig.2).
The evolution of the satellites depends  on the orbital parameters
and disk/orbit orientation as well. Encounters on the more eccentric orbits are
more damaging because the tidal field is stronger at small pericenters.
Prograde encounters are a lot more destructive compared to
retrograde ones  (Toomre \& Toomre 1972, Barnes 1988).
Large tidal tails appear only in prograde encounters  due to an
approximate resonance between internal and orbital motions
(Springel \& White 1998) and their extension is
considerably larger in LSB galaxies because of their shallower halo
density profile and larger disk scale-length (Fig. 2).

Stellar streams form which trace the orbital path of the satellite:
their patterns are long lived.
Tidal interactions have a profound influence on the morphology 
of the satellites, that evolve from disks to spheroids:
the degree of flattening of the remnants varies from case to
case, depending also on the disk/orbital plane orientation.    
The stellar remnants have final tidal radii
in the range  $6-10$ kpc. If we measure $D_{25}$, i.e.
the radius containing a surface brightness higher than $25$ mag
arcsec$^{-2}$, the size of our remnants would never exceed $3-4$ kpc (see
Fig.1).
Remnants of LSBs could even be missed by some optical surveys.
Remarkably, the projected density profile of the satellites steepens
but remains close to exponential, with final scale lengths smaller than
the initial ones.
The central surface brightness increases by up to $1$ mag arcsec$^{-2}$ 
for HSBs due to the angular momentum loss, while it
can decrease by about the same amount  for LSBs on the most eccentric
orbits (see Fig.1)
The central dispersion behavior seems to follow that of the
central surface density, increasing remarkably for HSB galaxies
(coarse grained phase space density decreases,  e.g. Hernquist et al.
1993). The final values are comparable with those observed in early-type
dwarfs in our Local Group ($10-30$ km/s).

\begin{figure}
\centerline{\epsfig{file=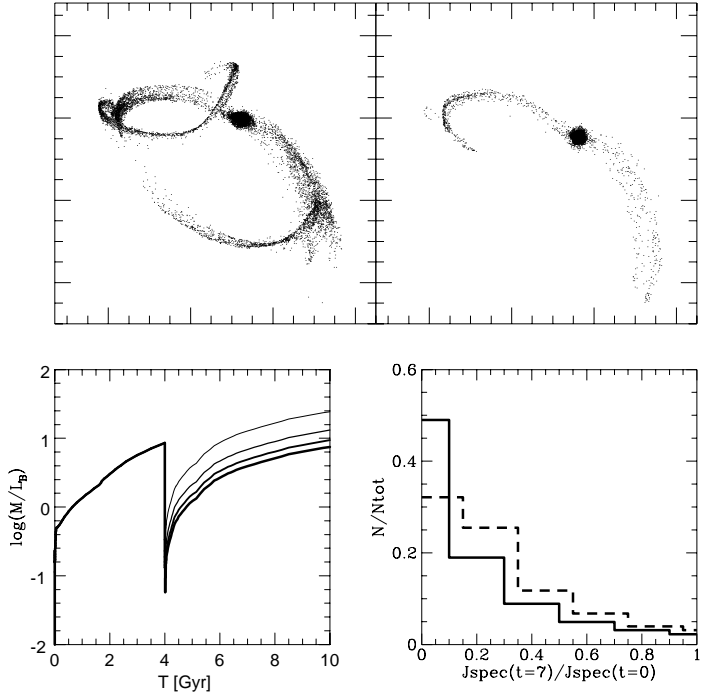,height =10cm}}
\caption{
Upper left  panel: disk particles of the LSB satellite after
a coplanar prograde encounter,
projected on the orbital plane. The box size is 390h$^{-1}$ kpc.
Upper right: same
for the HSB model.  Lower left panel: evolution of the M/L ratio (B band) 
for the LSB model assuming different burst strengths ($1,2,3,5
M_{\odot}$/yr),
with thicker lines for stronger bursts. 
Lower right:  distribution  of final vs initial  specific angular momentum
 for disk particles of   LSB (dashed) and HSB (solid) satellites. 
Only particles that end up in the remnant are shown.}
\end{figure}

To derive mass-to-light ratios of the satellites to be compared with
current observations, we have included a simple description of the
evolution of the stellar component in an LSB satellite using stellar
population synthesis models by Bruzual \& Charlot (1993)
with a Kennicutt IMF (Kennicutt  1994) and assuming a
metallicity $Z \sim 1/3 Z_{\odot}$, as typically
inferred for LSBs (Gerritsen \& de Blok 1999).  We suppose the galaxy
to
form at $z \sim 2$ and enter the virial radius of the primary at $z
\sim 1$, i.e. after 2 Gyr in our assumed cosmology.  

Its star
formation history is divided in three distinct phases: during the
first 2 Gyr we assume an exponentially declining star formation rate
(SFR) with a large time constant ($\tau = 10$ Gyr) and an amplitude
of $0.2 M_{\odot}/yr$, as suggested by observations and numerical
simulations (Gerritsen \& de Blok 1999). The satellite
enters the primary with the mass and luminosity of our N-Body model.
It then undergoes a central burst after the first
pericentric passage (at $t \sim 4$ Gyr) as a consequence of bar formation
and induced gas inflow (e.g. Lake et al. 1998).  Finally we assume star
formation to be truncated due to ram pressure and tidal stripping of gas,
leading to passive evolution until the present time.  The SFR during the
burst has amplitudes going from 1 to $\sim 5 M_{\odot}/yr$ and a duration
of $50$ Myr, as observed in blue star forming LSBs: a larger burst
would be quite inconsistent with the low gas
density and weak bar instability expected in an LSB galaxy.

At the present time the stellar mass-to-light ratio in the B band
has a lower limit of $\sim 7$ (Fig. 2).
A higher value, of the order of $11-12$, is obtained using solar
metallicity or smaller bursts. However, dSphs
have usually metallicity well less than solar (Grebel 1998).

Including the dark matter contribution, the final central mass-to-light
ratio of our LSB satellites  is  at least $10-15$.
However, mass-to-light ratios in dSphs are based on measures of the
line-of-sight velocity dispersion, from which the central density is
inferred.
The remnants of LSBs have extended tails that can project along the 
line of sight: we find that we can overestimate the central dispersion
by a factor of $\sim 2$ due to velocity gradients in the tails
(Platek \& Pryor 1995).
Thus, including also tidal effects we would measure central mass-to-light 
ratios in the range $10-40$ for the remnants of LSB satellites:
similar values are found for many dSphs, like Leo I
or Sagittarius A (Mateo 1998). Higher values
($ > 50$) are necessary to match those of Draco and Ursa Minor.
HSBs, on the contrary, are less affected by tides: 
the line-of-sight velocity dispersions are high at the end ($\sim$ 30
km/s) but reflect the velocity dispersion expected from virial
equilibrium.   
Low total mass-to-light ratios (of the order of $2-6$) 
are inferred  for dSphs/dEs like NGC205 (Mateo 1998).
These are  close to the values we obtain for $M_{total}/M_{stars}$
in the remnants of HSBs. Thus, if we want these satellites to
be ancestors of the brightest spheroidals, we need to suppose a more
prolonged star formation to maintain low stellar mass-to-light ratios.
Regions of recent star formation do exist in dSphs/dEs (Grebel 1998).

\section{Conclusions}    

This work  shows that {\it the long-lived interaction
between a satellite and the primary galaxy can drive a dramatic
morphological transformation between dwarf spirals to spheroidals}. 
After 2-3 pericenter passages LSB disk satellites resemble
currently observed dSphs galaxies (see Mateo 1998).  
HSBs, instead,  become more centrally concentrated and are likely 
the ancestors of more luminous dwarf ellipticals (dSphs/dEs).
The increase in concentration is  related to an interaction
driven  bar-instability which causes stars to lose substantial angular
momentum. 

The end products of interactions have lost plenty of dark matter
during their evolution. Stars are no more a secondary mass component at
the end, especially in the case of LSBs. 
However, a combination of fading of the stellar component and inflated
velocity dispersions due to projection of tidal tails can produce
M/L as high as those of many dSphs in 
the Local Group (Mateo 1998), while Draco and Ursa Minor
need further investigation.
A more prolonged star formation 
is requested to explain the observed M/L ratios of brighter
dSphs/dEs. 
                   
It's tempting to associate our suggested picture for the evolution of 
galaxy satellites with the observed population
of  blue-compact galaxies at intermediate and high-redshift (Guzm\'an et
al. 1997). 
%These objects have average observed circular velocities
%of $65$ km/s and are in some cases companions of larger spirals (***FG Guzm\'a%n, private communication). 
We propose that those galaxies are disk-satellites  undergoing morphological
evolution and interaction induced star formation.

Moore et al. (1998, 1999) has shown that fast encounters between
galaxies in clusters ("galaxy harassment") can drive their
morphological evolution from disks to spheroids.  This work shows  
that a similar scenario applies for large galaxy satellites, in a regime
where the ratio of relative/internal velocities is smaller.  These
results combine to show that tidal interactions provide a general, all
purpose mechanism to evolve galaxies along the Hubble sequence. This is an 
alternative to the classic merger scenario and is likely to occur in a large
variety of environments throughout the history of the Universe.

\vskip 0.5truecm

{\bf References}
 
\vskip 0.5truecm

Barnes J.E. 1988, {\it Ap.J.}, {\bf 331}, 699 

%Baugh, C.M., Cole, S., \& Frenk, C.S., 1996, {\it MNRAS}, {\bf 287}, L27 

Bruzual G. \& Charlot S. 1993, {\it Ap.J.} {\bf 405}, 538

Cole S., Lacey C. G., Baugh C. M.  \& Frenk C. 1999, in preparation

Colpi M., Mayer L., Governato F. \& Quinn, T. 1999, in preparation

de Blok W.J.G. \& McGaugh S.S. 1997, {\it M.N.R.A.S.}, 290,533

Dressler A., Oemler, A., Couch, W.J., Smail  I., Ellis R.S, Barger 

A., Butcher H., Poggianti B.M., Sharples R.M. 1998 {\it Ap.J.} {\bf 409},
577

Ghigna S., Moore B., Governato F., Lake G., Quinn T. \&

Stadel J. 1998, {\it M.N.R.A.S}, {\bf 299}, 515

Gerritsen J.P.E. \& de Blok W.J.G. 1999 , {\it A \& A},
{\bf 342}, 655

Grebel E.K., 1998, {\it astro-ph/9806191}

Guzm\'an R., Gallego J., Koo D.C., Phillips A.C., Lowenthal J.D., 

Faber S.M., Illingworth G.D., Vogt N.P. 1997, {\it Ap.J.}, {\bf 489}, 559

Hernquist L. 1993, {\it Ap.J.Suppl.}, {\bf 86}, 389

Hernquist L. Spergel D.N. Heyl J.S. 1993, {\it Ap.J.}, {\bf
416},415

Ibata R.A. \& Lewis, G.F. 1998, {\it Ap.J.}, {\bf 500}, 575

Kennicutt R.C. \& Tamblyn P. 1994, {\it Ap.J.}, {\bf 435}, 22

Lacey C. \& Ostriker J.P. 1985, {\it Ap.J.}, {\bf 176}, 1

Lake G., Katz N. \& Moore, B. 1998, {\it Ap.J.}, {\bf 495}, 152

Mateo M. 1998, {\it astro-ph/98100070}

Mihos  J.C., McCaugh S.S. \& De Blok W.J.G. 1997, {\it ApJL}, {\bf 477},L79

Moore B., Katz N., Lake G., Dressler A. \& Oemler A.  1996, {\it

Nature}, {\bf 379}, 613 

Moore B., Lake G. \& Katz N. 1998, {\it Ap.J.}, {\bf 435}, 139

Moore  B.,  Lake  G. , Quinn  T. \& Stadel  J. 1999, {\it MNRAS}, in press

Persic M. \& Salucci P. 1997, {\it Dark and visible matter in galaxies 

ASP ConferenceSeries}, 117 ed. M.Persic P.Salucci

Platek, S. \& Pryor, S. 1995, {\it Astron. Journal}, {\bf 109},
1071

Springel V. \& White S.D.M. 1998, {\it astro-ph/9807320}

Velazquez H. \& White S.D.M. 1998,{\it astro-ph/9809412}

Toomre \& Toomre 1972,  {\it Ap.J.}, 178, 623

%White  S.D.M. \& Frenk  C.S. 1991, {\it Ap.J.}, {\bf 379}, 25

Zaritsky D., Smith R., Frenk C. \& White S.D.M. 1993, {\it Ap.J.},

{\bf405}, 464

Zwaan  M.A., Van der Hulst J.M., de Blok  W.J.G. \& Mc Gaugh

S.S. 1995, {\it M.N.R.A.S.}, {\bf 273}, L35

\end{article}

\end{document}